\documentstyle[aps,prb]{revtex} \def\narrowtext{} \tighten \twocolumn

\begin{document}

\draft

\title{Direct observation of particle-hole mixing in the superconducting state
by angle-resolved photoemission}

\author{J.C. Campuzano$^{1,2}$, H. Ding$^{1,2}$, M.R. Norman$^1$, 
M. Randeria$^{1,3}$, A.F. Bellman$^{1,6}$, T. Yokoya$^4$,  
T. Takahashi$^4$, H. Katayama-Yoshida$^4$, 
T. Mochiku$^5$, K. Kadowaki$^5$}

\address{(1) Materials Science Division, Argonne National Laboratory, 
             Argonne, IL 60439,\\
         (2) Department of Physics, University of Illinois at Chicago, 
             Chicago, IL 60607,\\
         (3) Tata Institute of Fundamental Research, Bombay 400005, India \\
         (4) Department of Physics, Tohoku University, 980 Sendai, Japan,\\
         (5) National Research Institute for Metals, Sengen, Tsukuba, 
             Ibaraki 305, Japan,\\
         (6) Dipartimento di Fisica, Universit{\'a} di Milano, 
             20133 Milano, Italy \\
         }
\address{%
\begin{minipage}[t]{6.0in}
\begin{abstract}
Particle-hole (p-h) mixing is a fundamental consequence of
the existence of a pair condensate.
We present direct experimental evidence for p-h
mixing in the angle-resolved photoemission (ARPES)
spectra in the superconducting state of Bi$_2$Sr$_2$CaCu$_2$O$_{8+\delta}$.
In addition to its pedagogical importance, this establishes
unambiguously that the gap observed in ARPES is associated
with superconductivity.
\typeout{polish abstract}
\end{abstract}
\end{minipage}}

\maketitle
\narrowtext

A fundamental consequence of the existence of a pair condensate in
superconductors is the mixing of particles and holes in the neighborhood 
of the Fermi energy, $E_F$ \cite{SCHRIEFFER}. 
While there are innumerable experiments on superconductors
which give indirect evidence for particle-hole (p-h) mixing, 
it would be very nice, from a pedagogical point of view at least,
to have direct experimental evidence for such mixing.
Among the conventional probes
of superconductivity, Andreev reflection perhaps comes closest
to providing a direct test of such mixing, although this
is not (usually) a momentum-resolved probe.
Here we show that angle-resolved photoemission spectroscopy (ARPES)
reveals p-h mixing in the most direct and dramatic way: by the
appearance of states that do not exist above $T_c$.

We have recently provided strong evidence \cite{SUMRULE}, by studying
sum rules, that the impulse approximation is valid in ultraviolet
photoemission studies of the high $T_c$ materials, 
and that the experimental energy distribution curves
(EDCs) can be interpreted \cite{BACKGROUND}
in terms of the one-particle spectral function $A({\bf k},{\omega})$.
More specifically, since ARPES measures only occupied states,
the intensity is proportional to 
the Fermi function $f(\omega)$ times $A({\bf k},{\omega})$.
Using this we shall study p-h mixing in the superconducting state
of Bi$_2$Sr$_2$CaCu$_2$O$_{8+\delta}$ (Bi2212).

To set the stage for the experimental results it may
be useful to recall p-h mixing in the BCS framework
(even though there are aspects of the data
which are dominated by many body effects beyond
weak coupling BCS theory).
The BCS spectral function is given by \cite{SCHRIEFFER}
\begin{equation}
\pi A({\bf k},\omega)=
\frac{u_{\bf k}^2 \Gamma}{(\omega-E_{\bf k})^2+\Gamma^2} +
\frac{v_{\bf k}^2 \Gamma}{(\omega+E_{\bf k})^2+\Gamma^2}
\end{equation}
where the coherence factors are 
$v_{\bf k}^2=1-u_{\bf k}^2={1\over2}(1-{\epsilon}_{\bf k}/E_{\bf k})$
and $\Gamma$ is a phenomenological linewidth. 
The normal state energy ${\epsilon}_{\bf k}$ is measured from  $E_F$  
and the Bogoliubov quasiparticle energy is 
$E_{\bf k}=\sqrt{\epsilon_{\bf k}^2+\vert\Delta({\bf k})\vert^2}$, 
where $\Delta({\bf k})$ is the gap function. 
Note that only the second term in Eq.~1, with the $v_{\bf k}$-coefficient,
would be expected to make a significant contribution to the EDCs
at low temperatures.  

In the normal state above $T_c$,
the peak of $A({\bf k},{\omega})$ is at $\omega = \epsilon_{\bf k}$
as can be seen by setting $\Delta = 0$ in Eq.~1.
We would thus expect to see in ARPES
a spectral peak which disperses through zero binding energy as 
${\bf k}$ goes through ${\bf k}_F$ (the Fermi surface).
In the superconducting (SC) state, the spectrum changes from
$\epsilon_{\bf k}$ to $E_{\bf k}$ (Fig.~1).
As ${\bf k}$ approaches the Fermi surface
the spectral peak shifts towards lower binding energy,
but no longer crosses $E_F$.
Precisely at ${\bf k}_F$ the peak is at 
$\omega = |\Delta({\bf k}_F)|$, which is the closest
it gets to $E_F$. This is the manifestation of the gap
in ARPES \cite{OLSON,SHEN,DING}. 
Further, as ${\bf k}$ goes beyond ${\bf k}_F$,
in the region of states which were unoccupied above $T_c$,
the spectral peak {\it disperses back}, receding away from $E_f$, 
although with a decreasing intensity (Fig.~1). This is
the signature of particle-hole mixing. 

In the remainder of this paper we will present
high resolution ARPES data 
on the high temperature superconductor Bi2212. 
We shall find that there is clear 
experimental evidence for the anomalous dispersion described above
indicative of p-h mixing. We emphasize that this
is the only way (known to us) of asserting that the
gap seen by ARPES is due to superconductivity rather than
of some other origin, e.g., charge- or  spin-density wave formation.
Finally, we shall comment on features in
the experimental spectra which are determined by
many body effects and go beyond the simple BCS expression of Eq.~1.

The results presented below were obtained on the very high quality 
single crystals ($T_c = 87$K) which were used in our previous Bi2212
studies \cite{SUMRULE,DING,DING1}. 
The measurements were carried out at the University of Wisconsin 
Synchrotron Radiation Center, using a high resolution 4-meter normal incidence 
monochromator with a resolving power of $10^4$ at $10^{11}$ photons/s. 
Details about the samples
and the experimental procedure may be found in Ref.~\onlinecite{DING}.
Even though the momentum window of our 
spectrometer has a diameter of 0.074 $\AA^{-1}$ at 22 eV photon energy, in this 
experiment, data were taken at momentum intervals one fourth this value
because the spectral peak exhibits a sizable dispersion in the 
energy and momentum interval of interest to this experiment. 
As can be seen from the data, the momentum window of the 
spectrometer does not obscure the dispersion of the spectral peak. 

In order to best see p-h mixing one must have a large gap,
so that it is better to be near the $\bar{M}Y$ 
Fermi surface (FS) crossing \cite{SHEN,DING}.
(Our notation is $\Gamma = (0,0)$, ${\bar M}= (\pi,0)$
and $Y=(\pi,\pi)$, where $\Gamma{\bar M}$ is 
along the Cu-O bond direction.) 
However, the dispersion is very flat in the neighborhood
of $\bar{M}$, which makes it hard to establish the 
bending back of the spectral peaks.
On the other hand, while there is significant dispersion
in the diagonal $\Gamma Y$ direction, 
the gap is very small.  As a compromise,
the data in Fig.~2 are taken along a series of points in
momentum space along a path parallel to $\bar{M}Y$ beginning
about 0.7 of the way from $\Gamma$ to $\bar{M}$.

First we discuss the normal state ($T = 95$K) data shown in 
Fig.~2b. Only by contrasting the SC state data with the normal
state can one establish p-h mixing.
Note that the spectral features are very broad (non-Lorentzian) 
and asymmetric.  The large linewidth is due to many-body effects
in the spectral function. The asymmetry, at least in part, comes
from the fact that the peak of the EDC corresponds 
to that of $f(\omega)A({\bf k},\omega)$, and the Fermi function cuts off
what would have been the peak of the spectral function $A({\bf k},\omega)$.
Thus some care is needed in identifying the Fermi surface ${\bf k} = {\bf k}_F$,
since it is not a trivial matter to locate the peak of $A({\bf k},\omega)$ at
zero binding energy.  

To determine FS location, we 
use the sum rule \cite{SUMRULE} relating the energy-integrated
ARPES intensity to the momentum distribution $n({\bf k})$, shown as points in 
Fig.~3a (determined by integrating the normal state data over the range plotted
in Fig.~2b). As a background, we used the bottom EDC
(0.88) in Fig.~2b. The error bars are determined mostly by the background 
subtraction procedure. 
We then look at the momentum derivative of this integrated
intensity, and identify ${\bf k}_F$
from a peak in $\nabla_{{\bf k}}n({\bf k})$ \cite{FS_DERIVATIVE},
a plot of which is shown as a line in Fig.~3a. 
From this we see that the EDC labeled 0.58 corresponds to
${\bf k} = {\bf k}_F$. We emphasize that the small peaks seen 
in the EDCs
for ${\bf k}$ beyond ${\bf k}_F$ on the unoccupied side, are {\it not}
the peaks of the corresponding spectral functions
$A({\bf k},\omega)$, which are presumably at positive binding energy.
The EDC peaks come from the Fermi function cutting off
$A({\bf k},\omega)$, as explained above. 

We next turn to the superconducting state ($T = 13$K) data in 
Fig.~2a at exactly the same set of ${\bf k}$ points as the normal state 
data in (b).
The SC state data are plotted on a smaller energy 
range over which there is a significant increase in 
the lifetime, i.e., a decrease in the spectral linewidth.   Note that the
linewidths are smaller than the peak positions relative to $E_F$, in complete
contrast to the normal state data. The peaks in the superconducting  
state have resolution-limited widths, that is, one has true quasiparticles in
the SC state.

There are three qualitative differences between the normal
and SC state data: 
(1) shift of the spectral feature to lower binding energy due
to the opening of a gap \cite{OLSON,SHEN,DING};
(2) change in lineshape due to electron-electron
interactions and gap formation;
and (3) the change in the dispersion due to p-h mixing.
We will briefly discuss the first two points and then turn to 
the third, which is the main focus of this paper.

First, we would like to demonstrate unambiguously that a gap is indeed
present.  There have been suggestions in the literature \cite{MIYAKE} that one
does not
see a real gap, since the peak positions of the normal and SC state data do not
shift relative to one another as would be expected based on the replacement of
$\epsilon_{\bf k}$ by $E_{\bf k}$.
To refute this argument, we show in Fig.~4 the spectrum at
${\bf k}_F$ above and below $T_c$ in comparison to a reference spectrum
of Pt (in electrical
contact with Bi2212) used to establish the Fermi level.  One clearly sees
that above $T_c$, the leading edge slopes of Bi2212 and Pt match.  Below
$T_c$, there is a clear shift of the leading edge of the Bi2212 spectrum to
lower binding energy.  The apparent match of the peak positions between
normal and SC state is just an artifact of the Fermi function cut-off in
the normal state which makes the normal state peak appear as if it is below
$E_F$.  We also note that
the minimum excitation energy in the SC state occurs 
at the the same point as the normal state ${\bf k}_F$ (Fig.~3).

We have previously discussed the changes in the lineshape
upon cooling through $T_c$ in Refs.~\onlinecite{SUMRULE,DING1}. 
We only note here that we do not observe orders of magnitude 
decrease in the linewidth below $T_c$, 
as microwave experiments \cite{HARDY}
and thermal conductivity \cite{ONG} 
show. This is because, as indicated above, the width of our SC state 
spectra is limited by the energy resolution 
of our spectrometer.

We now contrast the dispersion of the spectral peaks
in the normal and superconducting states of Fig.~2.
Above $T_c$ the peak of $A({\bf k},\omega)$ disperses towards the 
Fermi energy (zero binding energy) and crosses it, 
as described above in detail.
However, in the SC state the same 
peak is seen to approach the gap value, which
is the closest it gets to the Fermi energy, 
and then recede away from it, with decreasing
intensity. 

It should be emphasized that the peaks seen
in the SC state EDCs
for ${\bf k}$ beyond ${\bf k}_F$ {\it are}, in fact,
the peaks of the corresponding spectral functions
$A({\bf k},\omega)$, in contrast to the normal state.
The Fermi function is sharp at $13 K$,
with a width of order few meV, and the gap
has pushed the spectral function down to
lower binding energies, thus $f(\omega)$
does not, by itself, produce a peak by cutting off the spectral
function, as it did above $T_c$.

By comparing the curves for ${\bf k}$ beyond ${\bf k}_F$ in the two panels in
Fig.~2, one can see that below $T_c$  
there are occupied states
which did not exist in the normal state. 
This is a clear signature of particle-hole mixing.

To summarize this, we show in Fig.~3b the peak positions
of the SC state data versus ${\bf k}$ (white dots). The error bars are based 
on fits of the peak positions after background subtraction. From these fits 
to the data using
Eq.~1 with an appropriate integration over the momentum window and
convolutions with the energy resolution and Fermi function \cite{DING}, we have
been able to extract off the normal state dispersion (black dots) assuming a SC
gap of 33 meV (this gap value being extracted from the spectrum at ${\bf k}_F$).
The error bars are asymmetric to account for a possible (d-wave) variation of 
the gap with ${\bf k}$. This dispersion is consistent with the normal state
data of Fig.~2.
We argue that this is a more accurate method of determination of the normal
state dispersion than looking at normal state data, since the latter have poorly
defined peak positions due to the large linewidths in the normal state.\cite{LI}
Note that the flatness of the SC state dispersion near ${\bf k}_F$ is reproduced
by the fits, arguing that it is a consequence of the finite momentum window
rather than some anomalous behavior.

It should be pointed out that while the BCS expression of Eq.~1
gives qualitative insight into the observed spectra, the detailed
form of the lineshape, and its $T$-dependence, arises from many-body
physics which is not contained in the simple BCS expression.
An example of such a feature is the
unusual transfer of spectral weight from the
incoherent part of the spectral function to the coherent part
upon cooling through $T_c$ as seen in Fig.~2. This is most pronounced near
${\bf k}_F$ where the very broad normal state spectrum (consistent with the
absence of a true quasiparticle peak)
evolves into a SC state spectrum with a sharp resolution limited
peak (representing a coherent quasiparticle).  This is also
evident from the $T$-dependent spectra given
in Ref.~\onlinecite{SUMRULE}.

In conclusion, we have shown that particle-hole mixing 
in superconductors is directly observable 
by angle-resolved photoemission spectroscopy. 

This work was supported by the U. S. Dept. of Energy,
Basic Energy Sciences, under contract W-31-109-ENG-38.
The Synchrotron Radiation Center is supported by NSF grant
DMR-9212658.

\begin{figure}
\caption{Schematic dispersion in the normal (thin line) and 
superconducting (thick lines) states following BCS theory.
The thickness of the superconducting state lines indicate 
the spectral weight given by the BCS coherence factors ($v^2$ below $E_F$
and $u^2$ above).}
\label{fig1}
\end{figure}

\begin{figure}
\caption{Superconducting state (a) and normal state (b) EDCs for the same
Bi2212 sample for the set of ${\bf k}$-values (1/a units) which are shown at the
top.  Note the different energy ranges.}
\label{fig2}
\end{figure}

\begin{figure}
\caption{
(a) Integrated intensity versus ${\bf k}$ from the normal state data of
Fig.~2b (black dots), i.e., the momentum distribution, $n_{{\bf k}}$.
Its derivative is shown as the solid curve (arbitrary scale).  The
Fermi surface (${\bf k} = {\bf k}_F$) is identified by a peak in the derivative
which corresponds within resolution to where $n_{{\bf k}}$ is ${1\over2}$.
(b) SC state peak positions (white dots) and normal state dispersion (black
dots) versus ${\bf k}$.
Note the backbending of the SC state dispersion for ${\bf k}$ beyond
${\bf k}_F$ which is a clear indication of particle-hole mixing.}
\label{fig3}
\end{figure}

\begin{figure}
\caption{
SC (T=13K) and normal state (T=95K) Bi2212 spectra (solid curves) versus
reference Pt spectra (gray curves).  Note the clear presence of a gap
below $T_c$.}
\label{fig4}
\end{figure}


\begin{references}
\bibitem{LI} For energies larger than 3$\Delta$, the linewidth in the SC
state becomes comparable to that in the normal state.  Therefore, only for
energies smaller than 3$\Delta$ can the peak of the spectrum be
identified with the quasiparticle binding energy.
\end{references}
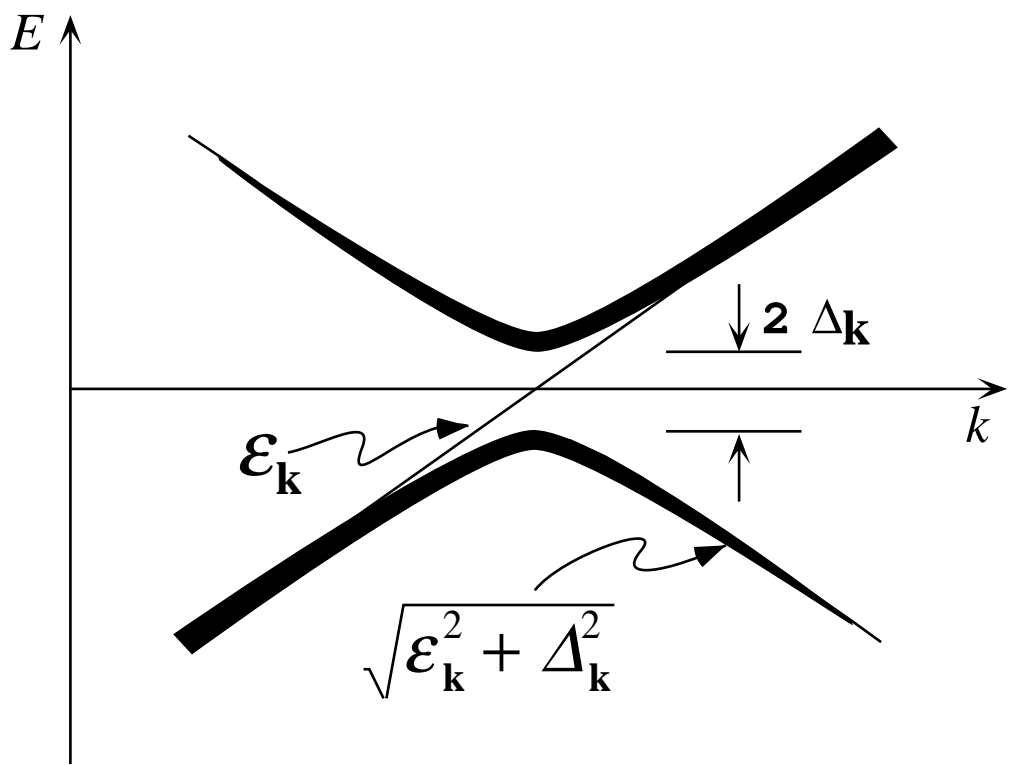
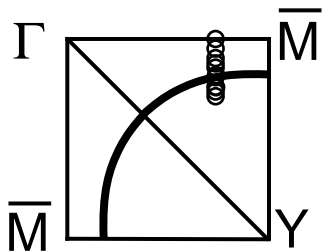
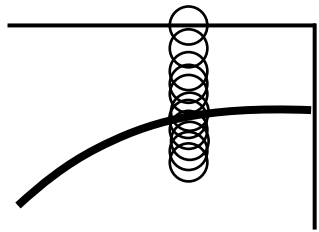
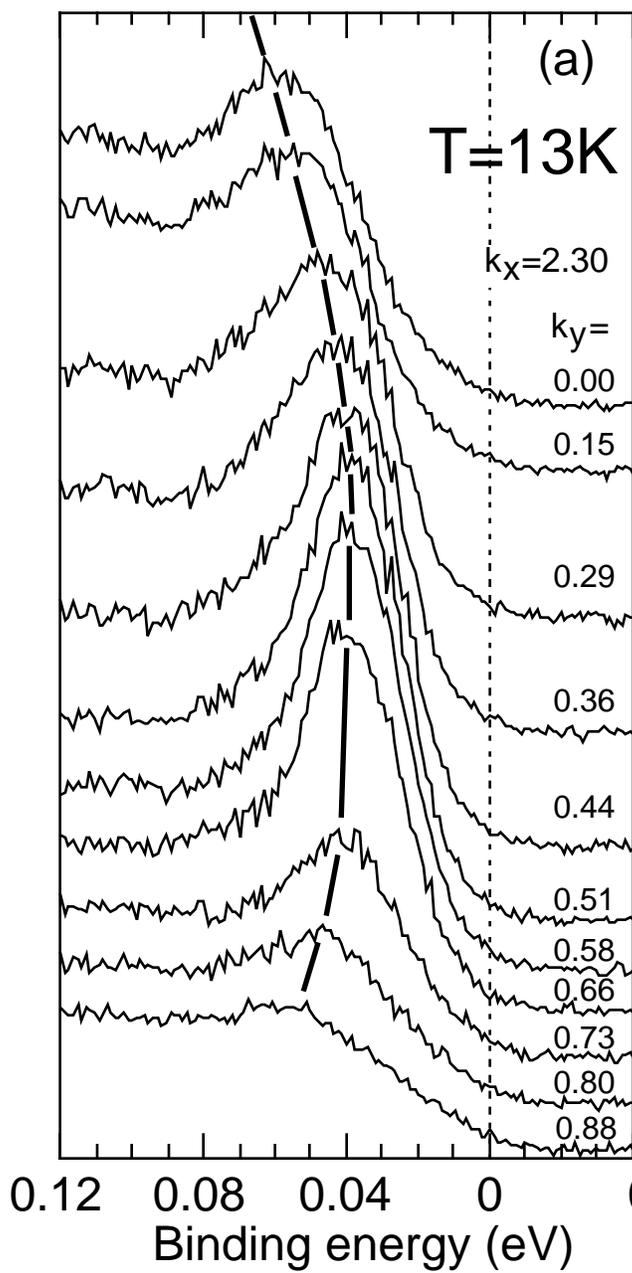
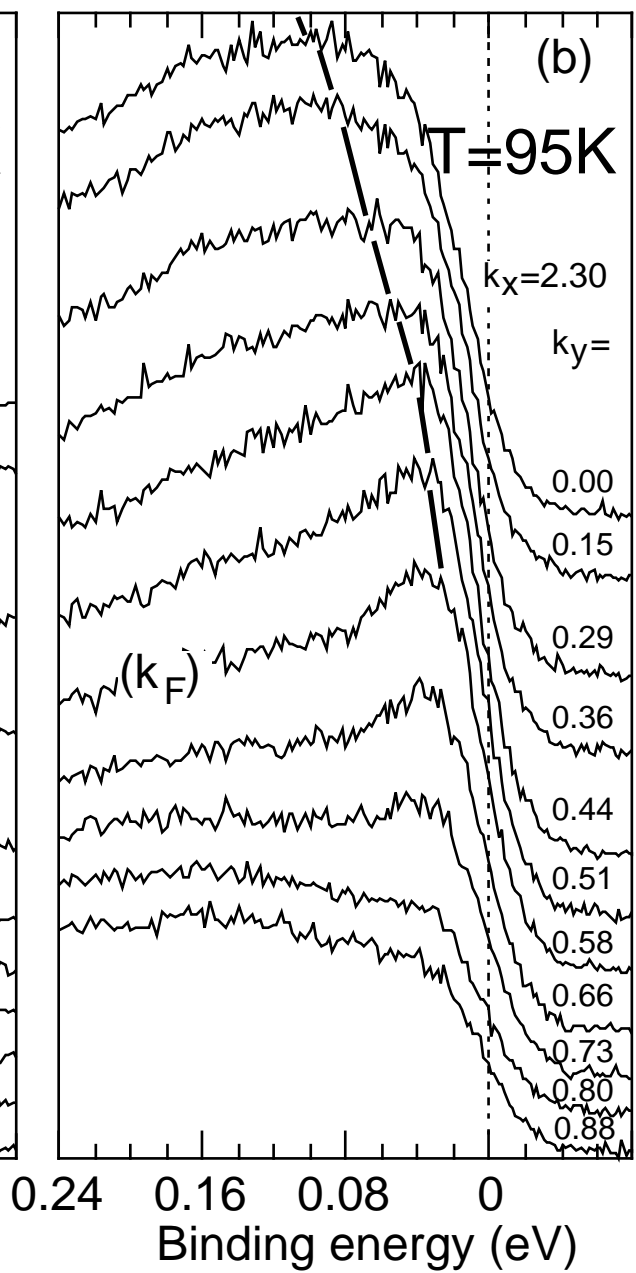
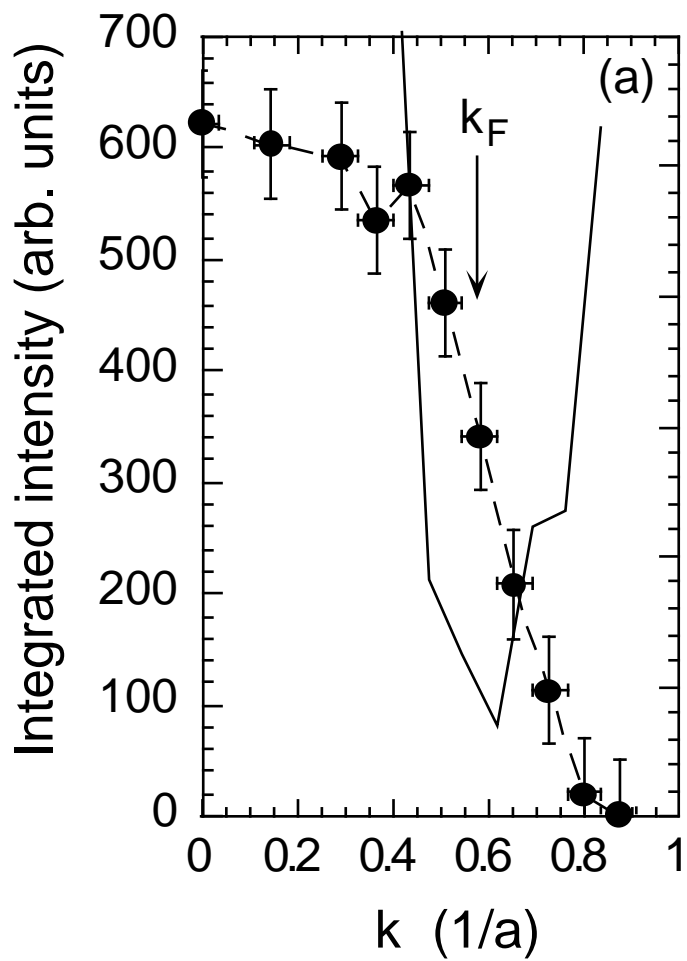
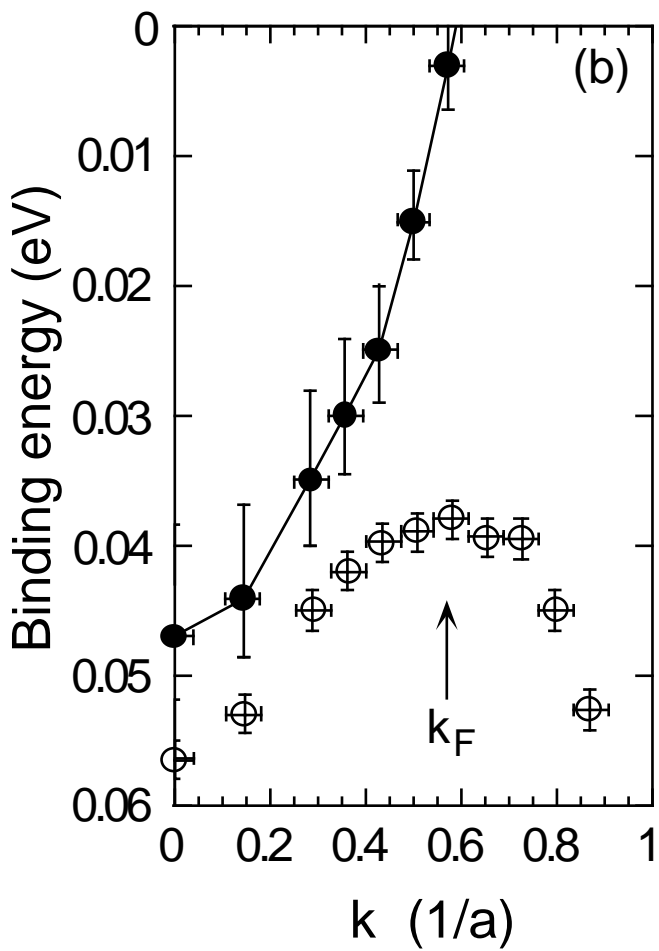
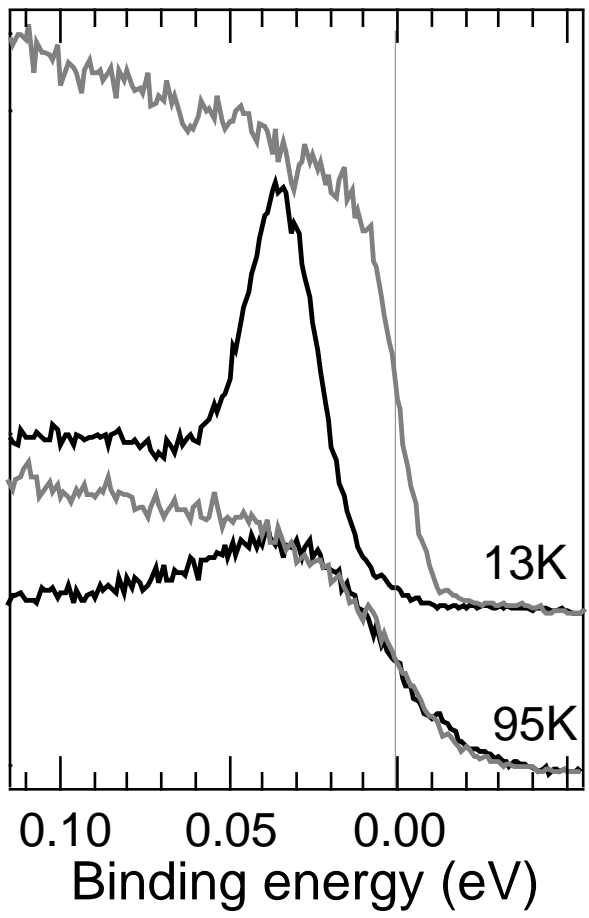
\end{document}